\documentclass[showpacs,preprint,aps]{revtex4}%

\usepackage{stmaryrd}
\usepackage{amstext}
\usepackage{amsfonts}
\usepackage{amsmath}
\usepackage{amssymb}
\usepackage{epsfig}
\usepackage{graphicx}%
\ifx\pdfoutput\relax\let\pdfoutput=\undefined\fi
\newcount\msipdfoutput
\ifx\pdfoutput\undefined\else
\ifcase\pdfoutput\else
\ifx\paperwidth\undefined\else
\ifdim\paperheight=0pt\relax\else\pdfpageheight\paperheight\fi
\ifdim\paperwidth=0pt\relax\else\pdfpagewidth\paperwidth\fi
\fi\fi\fi

\begin{document}

\title{Experimental self-testing of entangled states}

\author{Wen-Hao Zhang}
\affiliation{CAS Key Laboratory of Quantum Information, University of
Science and Technology of China, Hefei, 230026, China}
\affiliation{Synergetic Innovation Center of Quantum Information and Quantum Physics, University of Science and Technology of China, Hefei, Anhui 230026, China}

\author{Geng Chen$\footnote{email:chengeng@ustc.edu.cn}$}
\affiliation{CAS Key Laboratory of Quantum Information, University of
Science and Technology of China, Hefei, 230026, China}
\affiliation{Synergetic Innovation Center of Quantum Information and Quantum Physics, University of Science and Technology of China, Hefei, Anhui 230026, China}

\author{Xing-Xiang Peng}
\affiliation{CAS Key Laboratory of Quantum Information, University of
Science and Technology of China, Hefei, 230026, China}
\affiliation{Synergetic Innovation Center of Quantum Information and Quantum Physics, University of Science and Technology of China, Hefei, Anhui 230026, China}

\author{Xiao-Min Hu}
\affiliation{CAS Key Laboratory of Quantum Information, University of
Science and Technology of China, Hefei, 230026, China}
\affiliation{Synergetic Innovation Center of Quantum Information and Quantum Physics, University of Science and Technology of China, Hefei, Anhui 230026, China}

\author{Zhi-Bo Hou}
\affiliation{CAS Key Laboratory of Quantum Information, University of
Science and Technology of China, Hefei, 230026, China}
\affiliation{Synergetic Innovation Center of Quantum Information and Quantum Physics, University of Science and Technology of China, Hefei, Anhui 230026, China}

\author{Shang Yu}
\affiliation{CAS Key Laboratory of Quantum Information, University of
Science and Technology of China, Hefei, 230026, China}
\affiliation{Synergetic Innovation Center of Quantum Information and Quantum Physics, University of Science and Technology of China, Hefei, Anhui 230026, China}

\author{Xiang-Jun Ye}
\affiliation{CAS Key Laboratory of Quantum Information, University of
Science and Technology of China, Hefei, 230026, China}
\affiliation{Synergetic Innovation Center of Quantum Information and Quantum Physics, University of Science and Technology of China, Hefei, Anhui 230026, China}

\author{Zong-Quan Zou}
\affiliation{CAS Key Laboratory of Quantum Information, University of
Science and Technology of China, Hefei, 230026, China}
\affiliation{Synergetic Innovation Center of Quantum Information and Quantum Physics, University of Science and Technology of China, Hefei, Anhui 230026, China}

\author{Xiao-Ye Xu}
\affiliation{CAS Key Laboratory of Quantum Information, University of
Science and Technology of China, Hefei, 230026, China}
\affiliation{Synergetic Innovation Center of Quantum Information and Quantum Physics, University of Science and Technology of China, Hefei, Anhui 230026, China}

\author{Jian-Shun Tang}
\affiliation{CAS Key Laboratory of Quantum Information, University of
Science and Technology of China, Hefei, 230026, China}
\affiliation{Synergetic Innovation Center of Quantum Information and Quantum Physics, University of Science and Technology of China, Hefei, Anhui 230026, China}

\author{Jin-Shi Xu}
\affiliation{CAS Key Laboratory of Quantum Information, University of
Science and Technology of China, Hefei, 230026, China}
\affiliation{Synergetic Innovation Center of Quantum Information and Quantum Physics, University of Science and Technology of China, Hefei, Anhui 230026, China}

\author{Yong-Jian Han}
\affiliation{CAS Key Laboratory of Quantum Information, University of
Science and Technology of China, Hefei, 230026, China}
\affiliation{Synergetic Innovation Center of Quantum Information and Quantum Physics, University of Science and Technology of China, Hefei, Anhui 230026, China}

\author{Bi-Heng Liu}
\affiliation{CAS Key Laboratory of Quantum Information, University of
Science and Technology of China, Hefei, 230026, China}
\affiliation{Synergetic Innovation Center of Quantum Information and Quantum Physics, University of Science and Technology of China, Hefei, Anhui 230026, China}

\author{Chuan-Feng Li$\footnote{email:cfli@ustc.edu.cn}$}
\affiliation{CAS Key Laboratory of Quantum Information, University of
Science and Technology of China, Hefei, 230026, China}
\affiliation{Synergetic Innovation Center of Quantum Information and Quantum Physics, University of Science and Technology of China, Hefei, Anhui 230026, China}

\author{Guang-Can Guo}
\affiliation{CAS Key Laboratory of Quantum Information, University of
Science and Technology of China, Hefei, 230026, China}
\affiliation{Synergetic Innovation Center of Quantum Information and Quantum Physics, University of Science and Technology of China, Hefei, Anhui 230026, China}

\begin{abstract}
Quantum entanglement is the key resource for quantum information processing. Device-independent certification of entangled states is a long standing open question, which arouses the concept of self-testing. The central aim of self-testing is to certify the state and measurements of quantum systems without any knowledge of their inner workings, even when the used devices cannot be trusted. Specifically, utilizing Bell's theorem, it is possible to place a boundary on the singlet fidelity of entangled qubits. Here, beyond this rough estimation, we experimentally demonstrate a complete self-testing process for various pure bipartite entangled states up to four dimensions, by simply inspecting the correlations of the measurement outcomes. We show that this self-testing process can certify the exact form of entangled states with fidelities higher than 99.9$\%$ for all the investigated scenarios, which indicates the superior completeness and robustness of this method. Our work promotes self-testing as a practical tool for developing quantum techniques.
\end{abstract}

\maketitle

\section{Introduction}
In contrast to theoretical schemes of quantum information processing (QIP), where the imperfections of the involved devices are generally not taken into account, practically we often do not have sufficient knowledge of the internal physical structure, or the used devices cannot be trusted. The researches on this topic open a new realm of quantum science, namely, ``device-independent" science \cite{Acin1,Acin2,Masanes,Pironio,Lunghi1,Pal,Chen,Rabelo,Rabelo1}, in which no assumptions are made about the states under observation, the experimental measurement devices, or even the dimensionality of the Hilbert spaces where such elements are defined. In this case, the only way to study the system is to perform local measurements and analyze the statistical results. It seems to be an impossible task if we still want to identify the state and measurements under consideration. However, assuming quantum mechanics to be the underlying theory and within the no-signalling constraints, a purely classical user can still infer the degree of sharing entanglement due to the violation of the Bell inequalities \cite{Bell,Popescu1,Brunner1}, simply by querying the devices with classical inputs and observing the correlations in the classical outputs.

Such a device-independent certification of quantum systems is titled ``self-testing", which was first proposed by Mayers and Yao to certify the presence of a quantum state and the structure of a set of experimental measurement operators \cite{Mayers}. In the past decade, although other quantum features such as the dimension of the underlying Hilbert space \cite{Brunner,Gallego}, or the overlap between measurements \cite{McKague} can also be tested device-independently, most previous researches focus on the problem of
certifying the quantum entangled state that is shared between the devices \cite{Kaniewski,Moroder,Bardyn}. In particular, intensive studies have been devoted to the maximally entangled ``singlet" state, which is the cornerstone for QIP. Meanwhile, a large amount of progress has been made overall on the self-testing of other forms of entangled states. For example, Yang and Navascu$\acute{e}$ propose a complete self-testing certification of all partially
entangled pure two-qubit states \cite{Yang,Bamps}. In addition, the maximally entangled
pair of qutrits \cite{Salavrakos}, the partially entangled pair of qutrits that
violates maximally the CGLMP3 inequality \cite{Yang1,Acin3}, a small class
of higher-dimensional partially entangled pairs of qudits \cite{Coladangelo1}, and multi-partite entangled states \cite{Liang} and graph states \cite{Mckague1,Pal} are also shown to be self-testable. Another interesting application is the possibility of self-testing a quantum computation, which consists of
self-testing a quantum state and a sequence of operations applied to this state \cite{Magniez}.

High-dimensional bipartite entanglement is crucial for quantum-information tasks including teleportation using qudits \cite{Bennett}, generalized
dense coding \cite{Bennett1} and some quantum key distribution protocols \cite{Ekert}. As a result, self-testing high-dimensional bipartite
entanglement is a problem demanding prompt solution. Recently, A. Coladangelo $\emph{et al.}$ \cite{Coladangelo2} have provided a general method to self-test all pure bipartite entangled states by constructing explicit correlations, which can be achieved exclusively by measurements
on a unique quantum state (up to local transformations). In other words, this criterion can be as complete as one hopes for, rather than giving a bound on the singlet fidelity \cite{Navascues,Bancal}. Despite of considerable progresses made in self-testing theory, the relevant experimental work is still missing. In this work, by constructing a versatile bipartite entanglement source, we experimentally demonstrate a robust self-testing process of bipartite systems up to four dimensions. Our work indicates the practical feasibility of device-independent certification of entangled states by self-testing.

 \begin{figure}[htbp]
\centering
\includegraphics[width=6in]{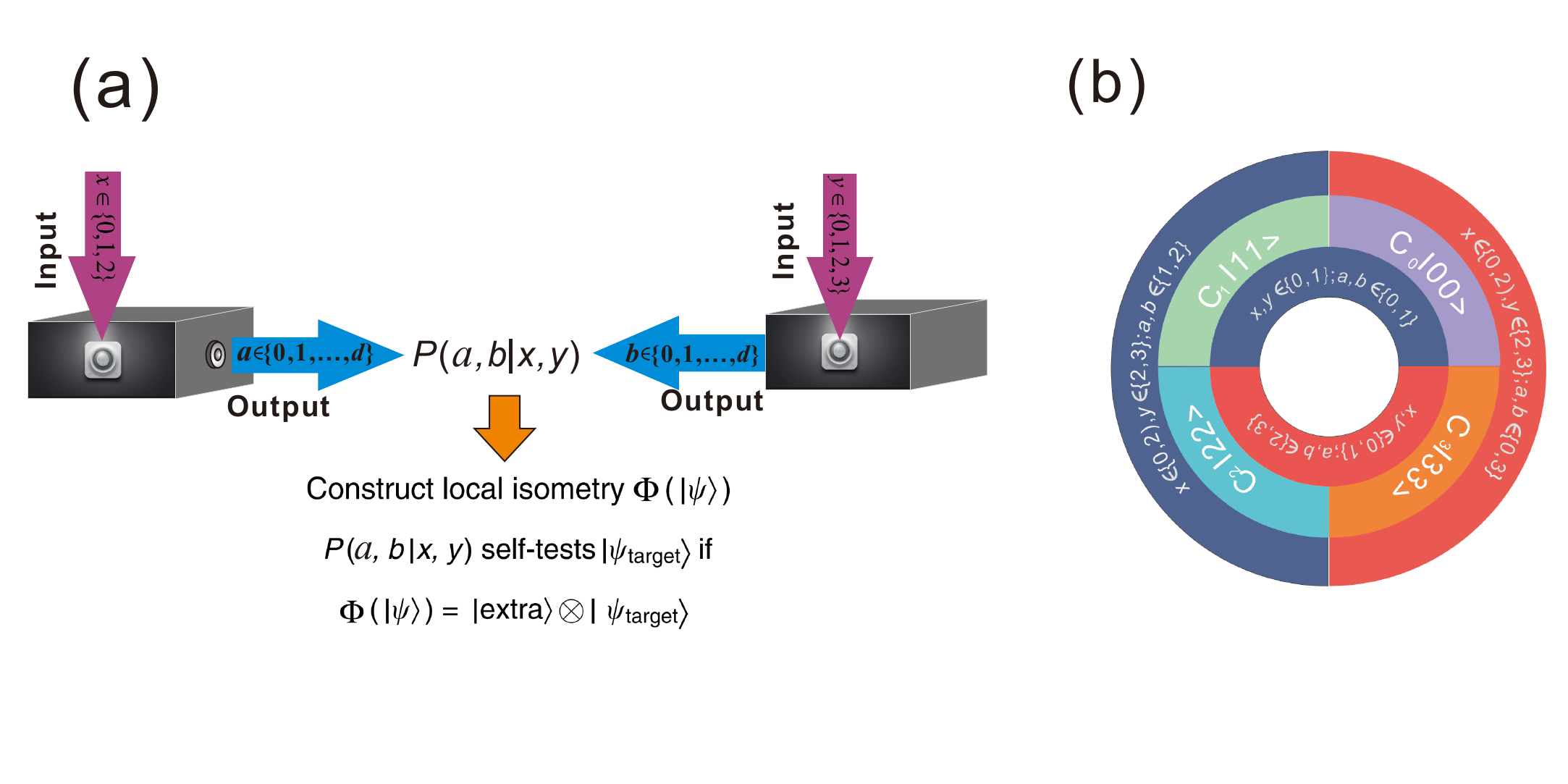}
\caption{\textbf{Self-testing proposals}. (a) The scheme to self-test an arbitrary pure bipartite entangled state: measurement inputs and outputs from a $[\{3, d\},\{4, d\}]$ Bell experiment are recorded, by which one can estimate the correlations of the Bell experiment and construct a local isometry $\Phi$. We conclude that the correlations self-test $|\psi_{target}\rangle$ on the condition of $\Phi(|\psi\rangle)=|auxiliary\rangle\otimes|\psi_{target}\rangle$. (b) The specific strategy to self-test a four-dimensional state in Eq. (\ref{highd}). This joint state is decomposed into four $2 \times 2$ blocks, and each block can be self-tested according to the measurement strategy illustrated, while the whole state can be self-tested by combining these results.}
\label{strategy}
\end{figure}

\section{Theoretical Framework}

The self-testing of arbitrary entangled two-qubit systems was resolved by Yang and Navascu$\acute{e}$ \cite{Yang}. In their work, they considered a scenario in which Alice and Bob share a pure two-qubit entangled state $|\varphi\rangle$, and one can record the probabilities in a $[\{2, 2\},\{2, 2\}]$ Bell
measurement, in which both Alice and Bob have two possible measurement settings with binary outcomes. From the measured correlations, it is possible to construct an isometry satisfying that follows equations
\begin{align}
& \Phi(|\varphi\rangle)=|\varphi_{target}(\theta)\rangle \\
& \Phi([\Pi_{A}\otimes\Pi_{B}]|\varphi\rangle)=[M_{A}\otimes M_{B}]|\varphi_{target}(\theta)\rangle.
\end{align}
$\Pi_{A}$, $M_{A}$ and $\Pi_{B}$, $M_{B}$ here stand for projective operators on the Alice and Bob sides, and $|\varphi_{target}(\theta)\rangle$ is a target state in the form of
\begin{equation}
\label{Bell}
|\varphi_{target}(\theta)\rangle=\cos\theta|00\rangle+\sin\theta|11\rangle.
\end{equation}

From this isometry, one can make a swap operation between the tested system and two ancillary qubits, after which one finds the ancillary qubits in certain target state $|\varphi_{target}(\theta)\rangle$. The corresponding procedure can be generalized to the relation $\Phi(|\varphi\rangle) = |extra\rangle \otimes |\varphi_{target}(\theta)\rangle$.
%This relation suggests any pure two-qubit state is equivalent to a certain partially entangled state $\cos\theta|00\rangle+\sin\theta|11\rangle$, up to local isometries.
%The remaining problem of composing a self-testing criterion now is how to identify measurable correlations from this relation.
It is further proved that the correlations embedded in this relation maximally and uniquely violate a particular family of Bell inequalities \cite{Acin4}, parametrized as
\begin{equation}
\label{Bell}
{\beta(\alpha)}=\alpha A_{0}+A_{0}(B_{0}+B_{1})+A_{1}(B_{0}-B_{1}),
\end{equation}
where $0 \leq \alpha \leq 2$ and the maximum quantum violation of it is $b(\alpha)=\sqrt{8+2\alpha^{2}}$.

Hereby, Yang and Navascu$\acute{e}$ gave the criterion to self-test all pure two-qubit states: When one observes a Bell correlation causing $b(\alpha_{0})-\langle\beta(\alpha_{0})\rangle=0$, the corresponding quantum state is equivalent, up to local isometries, to a certain entangled state $|\varphi_{target}(\theta)\rangle$ and $\theta$ is determined by the following equation:
\begin{equation}
\label{mapping}
\tan2\theta = \sqrt{\frac{4-(\alpha_{0})^{2}}{2(\alpha_{0})^{2}}}.
\end{equation}

Hence, it is proved that any pure two-qubit state can be self-tested, since its violation of Bell inequality Eq. (\ref{Bell}) is unique, up to
isometry. The authors also show that it is possible to generalize this method to bipartite high-dimensional maximally entangled states. However, it is still unclear whether arbitrary pure bipartite entangled state is self-testable.

Recently, A. Coladangelo $\emph{et al.}$ successfully addressed this long-standing open question by constructing explicit correlations built
on the framework outlined by Yang and Navascu$\acute{e}$s. The concrete process of this self-testing process is illustrated by Fig. 1. The uncharacterized devices are assigned to Alice and Bob, and they share a pure state as follows:
\begin{equation}
\label{highd}
|\psi\rangle = \sum_{i=1}^{d-1}c_{i}|ii\rangle.
\end{equation}

Initially they receive the inputs
$\emph{x}$ and $\emph{y}$ deciding their choice of measurement settings and return the outcomes $\emph{a}$
and $\emph{b}$, respectively. Consider a $[\{3, d\},\{4, d\}]$ Bell
scenario, in which
Alice has three possible measurement settings and Bob has
four, all of which have $\emph{d}$ possible outcomes. The measured statistics are recorded in the form of probabilities P$(a, b|x, y)$.
The central idea is to decompose $|\psi\rangle$ into two series of $2 \times 2$ blocks, and thus, each block can be self-tested through the two-qubit criterion described above \cite{Yang}. Referring to the four-dimensional states we test in the experiment, this decomposition can be implemented as shown in Fig. 1(b). Grouping P$(a, b|x, y)$ elements of which $x, y\in{0, 1}$, with $a, b\in{0, 1}$ and ${2, 3}$, one can certify $c_{0}|00\rangle+c_{1}|11\rangle$ and $c_{2}|22\rangle+c_{3}|33\rangle$ respectively. Similarly, using measurement settings $x\in{0, 2}$ and $y\in{2, 3}$, one can certify $c_{1}|11\rangle+c_{2}|22\rangle$ and $c_{0}|00\rangle+c_{3}|33\rangle$. Such a decomposition procedure can determine the relative ratio of the two modes in each block. The weight of each block in the joint state (6) can be further estimated by the total photon counting on this block, and eventually, the form of the tested state can be inferred.

 \begin{figure}[htbp]
\centering
\includegraphics[width=6in]{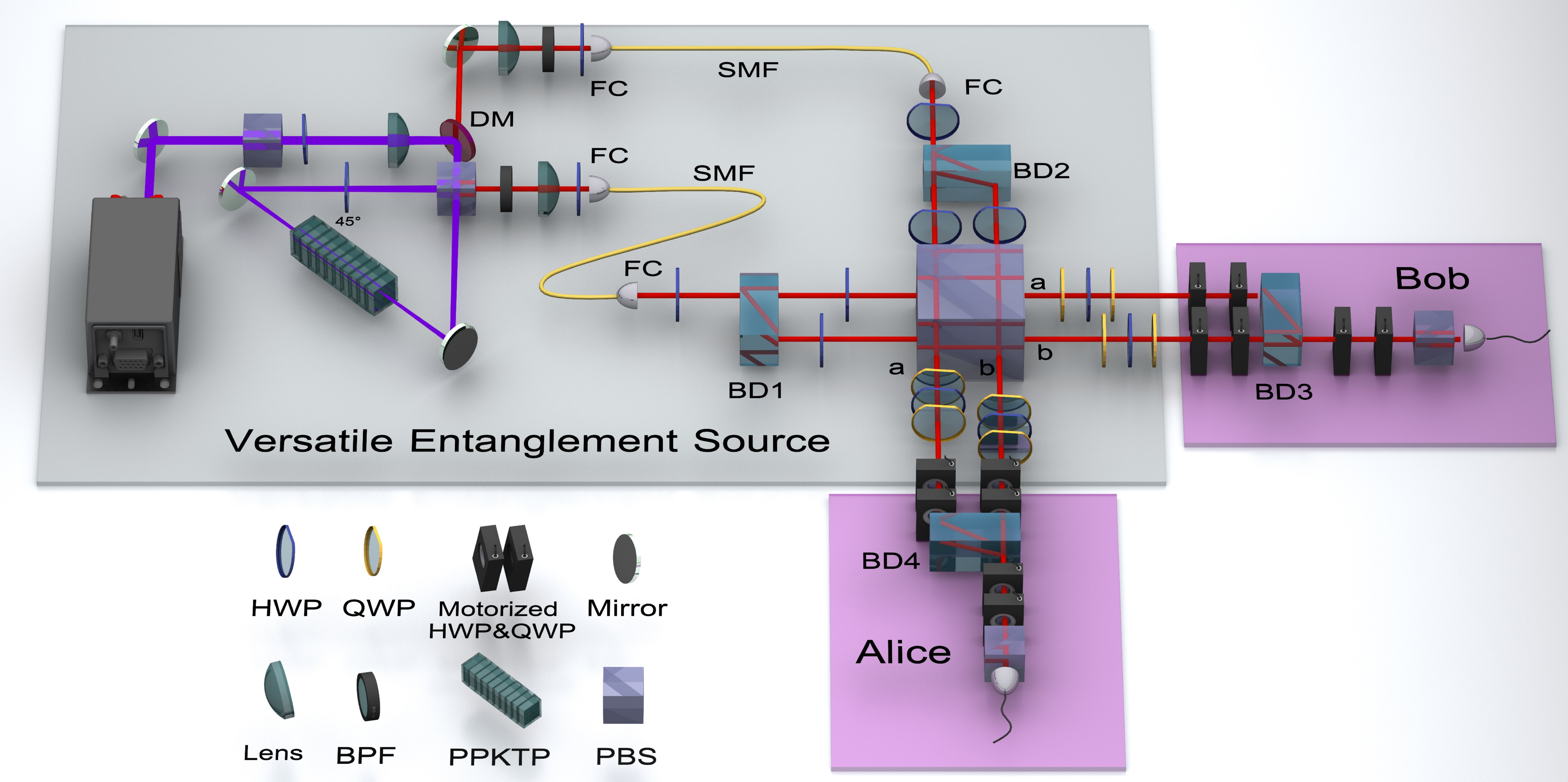}
\caption{\textbf{Experimental setup to self-test pure unknown bipartite systems.} The setup includes an entanglement source and a measurement apparatus for both the state tomograph and self-testing. In brief, entangled photon pairs are generated by pumping a periodically poled KTP (PPKTP) crystal in a polarization Sagnac interferometer (SI), and then they are encoded into polarization and path modes to produce high-dimensional entanglement. A hybrid measurement apparatus can perform arbitrary projective measurements on polarization and path modes. BD - beam displacer, HWP - half wave plate, QWP - quarter wave plate, DM - dichroic mirror, BPF - bandpass filter, PBS - polarized beam splitter, SMF -single mode fiber.}
\label{setup}
\end{figure}

\section{Experimental Results}

The criterion presented above is still a proof of ideal self-testing, which only considers perfect situations in which the correlations are exact.
Practically, however, the robustness to statistical noises and experimental imperfections is essential for self-testing proposals.
Due to the lack of theoretical analysis on the robustness bound, it is still unclear whether this complete criterion is feasible in the real world. In this work, we experimentally implement this complete self-testing proposal and study its precision and robustness by comparing the outcomes with state tomography results.
A versatile entangled photon-pair source is constructed which can generate pure bipartite states up to four dimensions, as shown in Fig. 2. Initially entangled photon pairs are generated by pumping a PPKTP crystal in a Sagnac interferometer (SI) \cite{Zeilinger}. Afterward, both photons are encoded into polarization and path modes, and therefore, the joint two-photon can be flexibly prepared into the product, two-qubit, two-qutrit and two-qudit entangled states (see Methods for details). Accordingly, the detecting apparatus is also properly designed in order to perform projective measurements on these two modes. Profiting from this setup, we can implement both self-testing and state tomography on the tested unknown states. We want to verify that the inferred states from the self-testing process are consistent with the tomography results, indicating the used self-testing criterion can indeed constitute a complete characterization for the quantum devices, and furthermore, of even greater importance, it is robust to realistic errors.

 \begin{figure}[htbp]
\centering
\includegraphics[width=6in]{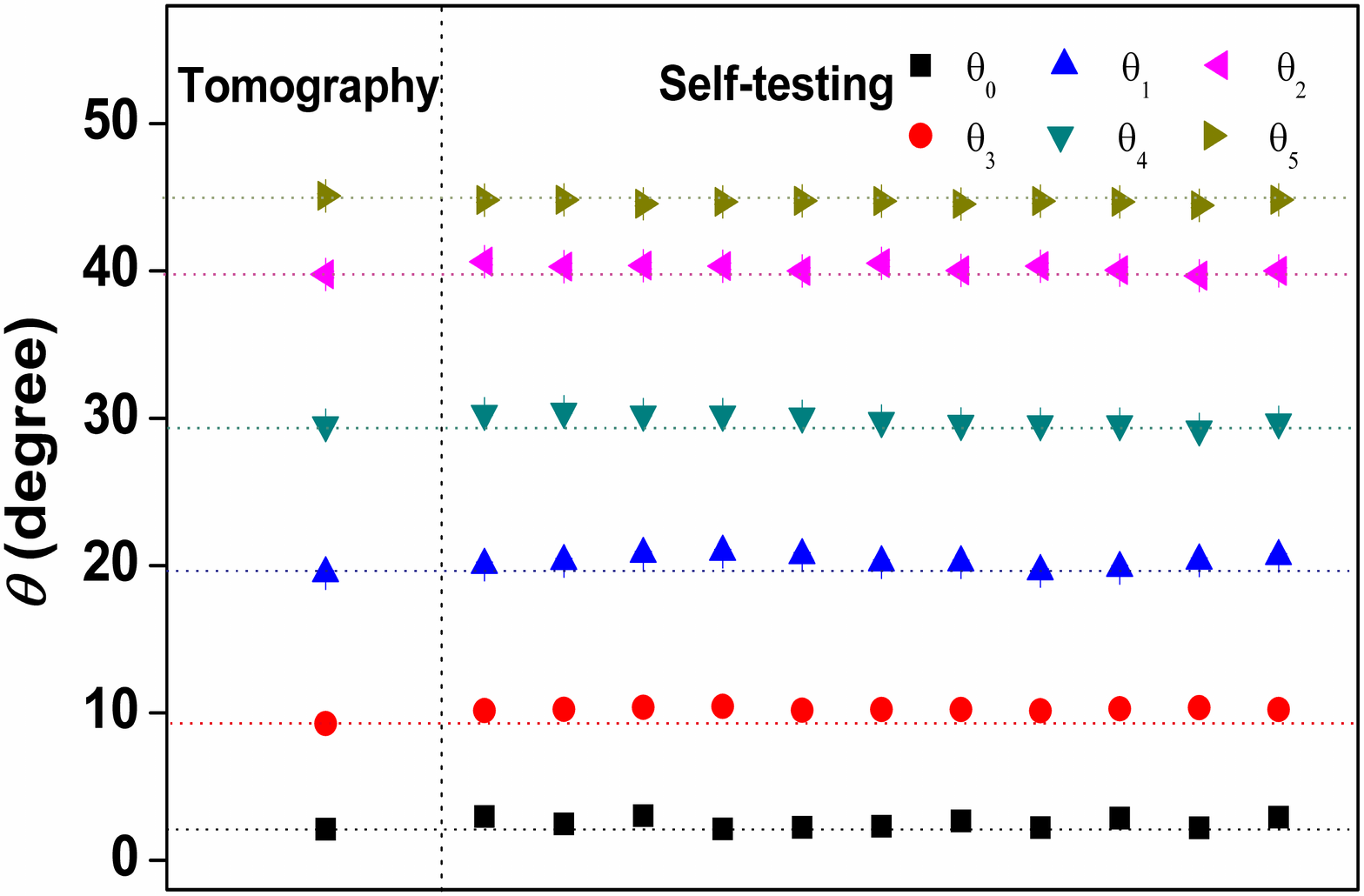}
\caption{\textbf{Self-testing results for six groups of two-qubit states.} For each group, we experimentally generated a two-qubit state $|\varphi_{j}\rangle$ ($j$=0,1,...,5), which only slightly deviates from the pure target state $|\varphi_{target}(\theta_{j})\rangle$. To each $|\varphi_{j}\rangle$, eleven unitary local operators are applied, resulting a series of equivalent transformations $|\varphi_{j(k)}\rangle (k=0,1,...,10)$. All of these states can be self-tested through a $[\{2, 2\},\{2, 2\}]$ Bell measurement. The first column, together with the dotted horizontal lines, indicate the tomography results of $\theta_{j}$, while the other eleven columns represent the self-testing outcomes of $\theta_{j}$ with different local operators. For each $|\varphi_{j(k)}\rangle$, the self-testing outcomes are very close to the tomography results.}
\label{results}
\end{figure}

 \begin{figure}[htbp]
\centering
\includegraphics[width=5in]{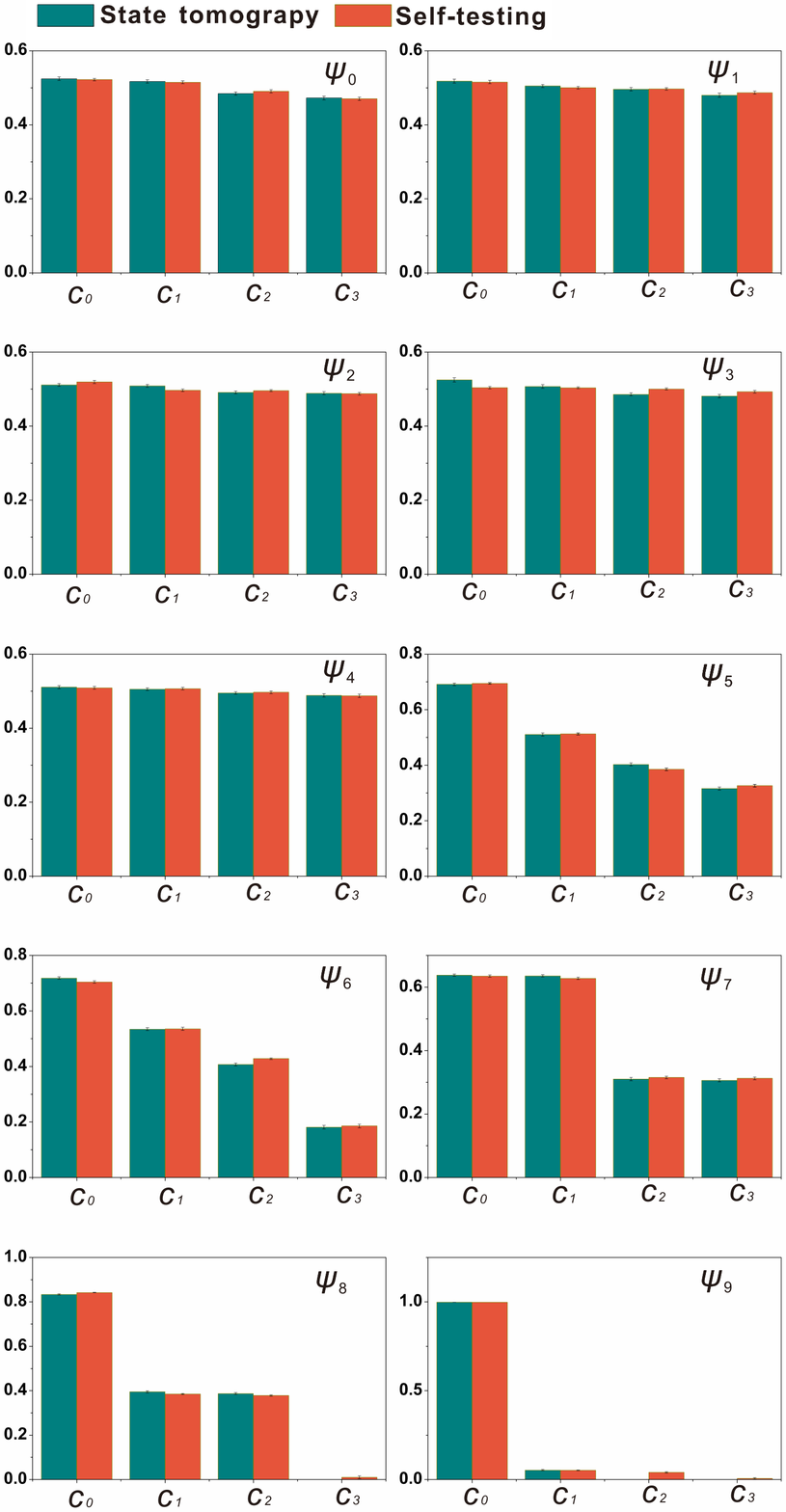}
\caption{\textbf{self-testing of high-dimensional bipartite states.} Through a $[\{3, d\},\{4, d\}]$ Bell experiment, totally ten pure states are self-tested. Approximately, $|\psi_{0}\rangle$ is a maximally entangled two-qudit state, and $|\psi_{1,2,3,4}\rangle$ are transformations of it through local unitary operations. $|\psi_{5,6,7}\rangle$ are partially entangled two-qudit states, and $|\psi_{8}\rangle$ is a partially entangled two-qutrit states, while $|\psi_{9}\rangle$ is expected to be a product state. For each state, both state tomography and self-testing results of the values of $c_{i}$ are shown. The agreement of these results indicates that these states can be precisely characterized by self-testing.}
\label{results}
\end{figure}
$\textbf{Two-qubit scenario.}$ As discussed above, the two-qubit self-testing criterion is the foundation for the entire theoretical frame. First, we test substantial two-qubit states through $[\{2, 2\},\{2, 2\}]$ Bell measurements.
In the experiment, we initially generate 6 two-qubit entangled states $|\varphi_{j}\rangle$ ($j$=0,1,...,5), each of which is in principle equivalent to the pure target state $|\varphi_{target}(\theta_{j})\rangle$. For each $|\varphi_{j}\rangle$, eleven different local operators are applied, which are randomly assigned by a program. Each operator leads to an unknown state $|\varphi_{j(k)}\rangle (k=0,1,...,10)$ that we want to self-test. For a certain $|\varphi_{j(k)}\rangle$, we perform Bell measurements, and the value of $\alpha_{0}$ is obtained when $b(\alpha_{0})-\langle\beta(\alpha_{0})\rangle$ approaches 0. The form of each tested state can then be identified by calculating the corresponding value of $\theta_{j}$ from Eq. (\ref{mapping}). In order to estimate the actual value of $\theta_{j}$, we measure the density matrix $\rho$ of $|\varphi_{j}\rangle$ through a state tomography process, and the actual value of $\theta_{j}$ is calculated as $\arctan(\sqrt{\frac{\langle 1|\rho|1\rangle}{\langle 0|\rho|0\rangle}})$. All of these results are shown in Fig. 3, with the first column showing the tomography results and other columns representing the self-testing outcomes for different $|\varphi_{j(k)}\rangle$. For all of the tested states, the self-testing outcomes are approximately identical to the corresponding tomography results, which clearly verifies the correctness of the applied self-testing criterion.

$\textbf{Two-qudit (qutrit) scenario.}$ For high-dimensional scenarios, we decompose the joint state into several $2 \times 2$ blocks, each of which can be self-tested similarly to the two-qubit case. The probabilities needed to self-test these blocks are measured according to the strategy shown in Fig. 1(b). When $d=4$, in the self-testing process, Alice has 12 possible projective measurements, and Bob has 16, which results in a total of 192 values of P$(a, b|x, y)$ (see Supplementary Info. for details). For each block, 16 values of P$(a, b|x, y)$ are recorded. As a result, 64 measurements are required to self-test the four decomposed blocks. The summation of the counts on each group of 16 projective measurements can be used as the weight of the corresponding block, and thus, the exact form of the joint state can be identified. For state tomography, both Alice and Bob have 16 projective measurements thus a complete reconstruction of the density matrix $\rho$ requires 256 measurements. The actual value of $|c_{i}|$ is calculated as $|c_{i}|=\sqrt{\langle i|\rho|i\rangle}$. We investigate totally 9 high-dimensional bipartite entangled states plus a product state (see Supplementary Info. for details), and the results are shown in Fig. 4. For all of the tested states, the inferred states from self-testing are in good accordance with the state tomography results. Therefore, it is proved that the utilized self-testing criterion can precisely certify an unknown pure bipartite state in a device-independent way.

$\textbf{Robustness to errors.}$
\begin{table}[!hbp]
\begin{tabular}{|c|c|c|c|c|c|c|c|c|c|c|c|}

\hline
$|\varphi_{j}\rangle$ & $|\varphi_{0}\rangle$ & $|\varphi_{1}\rangle$ & $|\varphi_{2}\rangle$ & $|\varphi_{3}\rangle$ & $|\varphi_{4}\rangle$ & $|\varphi_{5}\rangle$ \\
\hline
$P (\%)$ & 99.94 & 98.99 & 99.28 & 97.76 & 97.43 & 96.56 \\
\hline
$F_{S}(min) (\%)$ & 99.97 & 99.96 & 99.94 & 99.97 & 99.98 & 99.99
\\
\hline
$b(\alpha_{0})-\langle\beta(\alpha_{0})\rangle$ & 0.04282 & 0.0928 & 0.00691 & 0.05136 & 0.05324 & 0.04566
\\
\hline
\end{tabular}
\caption{Two-qubit self-testing errors in the presence of experimental imperfections. For each $|\varphi_j\rangle$, it is slightly mixed and its purity $P$ is smaller than 1. $F_{S}(min)$ represents the lowest self-testing fidelities to $|\varphi_{target}(\theta_{j})\rangle$ of eleven equivalent transformations from $|\varphi_j\rangle$, and the corresponding values of $b(\alpha_{0})-\langle\beta(\alpha_{0})\rangle$ are shown in the bottom line.}
\end{table}

\begin{table}[!hbp]
\begin{tabular}{|c|c|c|c|c|c|c|c|c|c|c|c|}

\hline
$|\psi_{j}\rangle$ & $|\psi_{0}\rangle$ & $|\psi_{1}\rangle$ & $|\psi_{2}\rangle$ & $|\psi_{3}\rangle$ & $|\psi_{4}\rangle$ & $|\psi_{5}\rangle$ & $|\psi_{6}\rangle$ & $|\psi_{7}\rangle$ & $|\psi_{8}\rangle$ & $|\psi_{9}\rangle$ \\
\hline
$P (\%)$ & 97.24 & 96.54 & 95.79 & 92.68 & 94.53 & 94.20 & 95.20 & 94.80 & 96.65 & 98.93 \\
\hline
$F_S (\%)$ & 99.99 & 99.99 & 99.99 & 99.96 & 99.99 & 99.98& 99.97 & 99.99 & 99.98 & 99.92
\\

\hline
\end{tabular}
\caption{The state purity ($P$) of the ten tested states in Fig. 4 and the corresponding self-testing fidelity ($F_S$) to expected pure target state in the form of Eq. (6), with $|c_{i}|=\sqrt{\langle i|\rho|i\rangle}$.}
\end{table}

In theory, self-testing statements of practically relevant robustness are significantly difficult to prove, and for many of the proposals the robustness is extremely weak. For the two-qubit self-testing criterion used here, Yang and Navascu$\acute{e}$s prove that when the measured maximum Bell violation slightly deviates from the theoretical value, the self-testing procedure can still return a quantum state that is close to $|\varphi_{target}(\theta_{j})\rangle$ \cite{Yang}. Therefore, it is reasonable to expect that the extended strategy for high-dimensional scenarios is also robust to small errors. The analytical robustness bound is still missing; nevertheless, we can study the robustness in a practical way.

The versatile entanglement source consists of sequential quantum interferences (See Method for details), as a result, the generated state suffers dephasing process and the final state becomes slightly mixed. The error in these states can be characterized by calculating the purity $P$ of the measured density matrix $\rho$, as $P=Re(tr(\rho^{2}))$.

For two-qubit scenarios, the practical robustness can be reflected by the errors in the self-testing outcomes as shown in Table I. For each experimentally generated state $|\varphi_{j}\rangle$, the state purity $P$ is smaller than 1 as shown in the second line of Table I. As a result, Eq. (\ref{Bell}) cannot reach the maximum violation and $b(\alpha_{0})-\langle\beta(\alpha_{0})\rangle$ is a nonzero value. In this case, the self-testing precision can be quantified by the fidelity of the inferred state to $|\varphi_{target}(\theta_{j})\rangle$. The lowest fidelity in each group is shown in the third line of Table I, in which the entries are all above 0.999 indicating a superior precision and robustness.

For high-dimensional scenarios, due to suffering imperfect Hong-Ou-Mandel interference (see Method sec. I for details), the purity of the generated states is even lower, as shown in the second line of Table II. However, we can still obtain satisfactory results and the self-testing fidelities are all very close to 1 as shown in the third line of Table II.

$\textbf{No-signalling constraints.}$
Device-independent certifications require no-signalling constraints \cite{Popescu} on the devices, which can be tested through the influence of the measurement of Alice (Bob) side on Bob (Alice) side \cite{Brunner1}. Concretely, no-signalling constraints require the following relations to be satisfied:
\begin{align}
& \sum_{b=0}^{d-1}P(a,b|x,y)=\sum_{b=0}^{d-1}P(a,b|x,y')\quad for\quad all\quad a,x,y,y' \\
& \sum_{a=0}^{d-1}P(a,b|x,y)=\sum_{a=0}^{d-1}P(a,b|x',y)\quad for\quad all\quad b,y,x,x'.
\end{align}
In experiment, all the 192 probabilities need to be recorded to verify above equations. In Supplementary Info. Sec. III we give the results when the test state is $|\psi_0\rangle$.

\section{Discussion}
There are several ways that a purely classical user can certify the quantum state of a system, among which the standard quantum tomography is the most widely used method. However, it depends on characterization of the degrees of freedom under study and the corresponding measurements, thereby, it becomes invalid if the devices cannot be trusted. Furthermore, when facing high-dimensional scenarios, an exponentially large number of projective measurements is required. Recently, Chapman $\emph{et al.}$ proposed a self-guided method which exhibits better robustness and precision for quantum state certification \cite{Chapman}. Unsurprisingly, all of these benefits rely on the ability to fully characterize the measurement apparatus. The presence of self-testing suggests that the state certification can also be implemented in a device-independent way. A substantial theoretical studies have focused on this new subject, while the relevant experimental studies are quite rare. Two main possible reasons for the lack of experiments could be the weak robustness and the trivial bounds in the previous theoretical proposals. Thanks to the consecutive efforts by Yang, Navascu$\acute{e}$, Coladangelo, Goh and Scarani, for the first time, a complete self-testing proposal has been raised, which is valid for all pure bipartite quantum states utilizing a $[\{3, d\},\{4, d\}]$ Bell measurement. We perform this method with our high-quality entanglement source and measurement apparatus. For all of the tested scenarios, the obtained results exhibit excellent precision and robustness. Due to this significant progress, self-testing can now be applied to realistic quantum technologies.

\section{Method}
\textbf{Versatile entanglement source.}
The versatile entanglement source consists of two main parts. One part is responsible for the generation of polarization entangled photon pairs, and the other part is in charge of entangling the two photons in both polarization and path modes to form high-dimensional states. In the first part, a 405 nm continuous-wave diode laser with polarization set by a half-wave plate is used to pump a 4 mm-long PPKTP crystal inside an SI to generate polarization-entangled photons. The photon pairs are in the state $\cos\theta|HH\rangle+\sin\theta|VV\rangle$ ($H$ and $V$ denote the horizontal and vertical polarized components, respectively) and $\theta$ is controlled by the pumping polarization. The visibility of the maximally entangled state is larger than 0.985. These photon pairs are then sent into the second part by two single mode fibers, and the polarization is maintained by two HWPs before and after each fiber. The path mode is then created by BD1 and BD2, which causes an approximately 4 mm separation between the $H$ and $V$ components at 810 nm. At this stage the path mode is entangled with the polarization mode, and therefore, the joint state remains two-dimensional. After BD1 and BD2, an HWP is inserted on each path and totally four HWPs are employed for state-preparing. With post-selection by Alice and Bob, two photon Hong-ou-Mandel interference at the crossing PBS can generate polarization entanglement on both path modes ($a$ and $b$). The visibility of this Hong-Ou-Mandel interference is one main obstacle of having high fidelity. Due to the bandwidth difference from the type II SPDC process, the best visibility we achieved here is $\sim 0.975$. The number of dimensions of the outcome states after the PBS can be selected by the alignments of the four state-preparing HWPs after BD1 and BD2. Concretely, when the four HWPs are set to be $0^{\circ}$ or $45^{\circ}$, the outcome is a path mode entangled state $\cos\theta|aa\rangle+\sin\theta|bb\rangle$. When HWPs rotate single $H$ or $V$ polarized photons to a superposition state $\alpha |H\rangle+\beta |V\rangle (\alpha,\beta \neq 0)$, the result is a two-qutrit state $|\psi\rangle (d=3) = c_{0}|00\rangle+c_{1}|11\rangle+c_{2}|22\rangle$. When both $H$ and $V$ polarized photons are rotated to be superposition states, we can finally obtain a two-qudit state $|\psi\rangle (d=4) = c_{0}|00\rangle+c_{1}|11\rangle+c_{2}|22\rangle+c_{3}|33\rangle$. All of the coefficients $c_{i}$ are decided by the rotating angle of the four HWPs and the value of $\theta$. The encoding rule uses $0,1,2,3$ to denote $|Ha\rangle,|Va\rangle,|Hb\rangle,|Vb\rangle$ components, respectively. In order to attain a larger state space, local unitary operations can be applied by the waveplate sets after the crossing PBS.

\textbf{Measurement apparatus.}
The measurement apparatus in this experiment is properly designed so as to perform both high-dimensional quantum tomograph and self-testing. On both the Alice and Bob sides, the measurement apparatuses have an array of QWP-HWP on both the $a$ and $b$ paths, which is in charge of making projective measurements on the polarization mode. A subsequent BD combines the two separated paths, and a QWP-HWP set with a PBS are responsible for the measurements on the path mode. All the waveplates are mounted in the programmable motorized rotation stages. It is obvious that the two pairs of BDs compose Mach-Zehnder like interferometers. Each BD used here is selected to be most compatible with its partner thus a high interference visibility of above 0.999 can be attained. For a four-dimensional state, we perform 256 joint projective measurements for state tomography and 64 for self-testing .

{\bf  Acknowledgments}

 This work was supported by the National Key Research and Development Program of China (Nos. 2017YFA0304100, 2016YFA0302700), National Natural Science Foundation of China (Grant Nos. 61327901, 11774335, 91536219), Key Research Program of Frontier Sciences, CAS (No. QYZDY-SSW-SLH003), the Fundamental Research Funds for the Central Universities (Grant No. WK2030020019, WK2470000026).
%\clearpage
{xx}

\end{document}